\documentclass[11pt]{iopart}

\usepackage{iopams}
\usepackage[utf8]{inputenc}

\usepackage{amsfonts, amssymb, latexsym}
\usepackage{graphicx}
\usepackage{rotating}
\usepackage[sort&compress,numbers]{natbib}
\usepackage{xspace}

\usepackage{xcolor}

\def\Lie{\mathcal{L}}
\def\G{\mathcal{G}}
\def\H{\mathcal{H}}
\def\M{\mathcal{M}}

\def\dif{\textrm{d}}

\newcommand{\Ord}[2]{\mathcal O \left(#1\right)^{#2}}

\begin{document}

\title{Black hole hair formation  in shift-symmetric generalised scalar-tensor gravity}

\author{Robert Benkel$^{\dagger}$, Thomas P. Sotiriou$^{\dagger,\sharp}$, Helvi Witek$^{\dagger,\star}$}

\address{$^{\dagger}$ School of Mathematical Sciences, University of Nottingham,
University Park, Nottingham, NG7 2RD, UK}
\address{$^{\sharp}$ School of Physics and Astronomy, University of Nottingham,
University Park, Nottingham, NG7 2RD, UK}
\address{$^{\star}$ Departament de F\'{i}sica Qu\`{a}ntica i Astrof\'{i}sica
\& Institut de Ci\`{e}ncies del Cosmos (ICCUB),
Universitat de Barcelona, Mart\'{i} i Franqu\`{e}s 1, E-08028 Barcelona,
Spain}

\eads{
\mailto{robert.benkel@nottingham.ac.uk},
\mailto{thomas.sotiriou@nottingham.ac.uk},
\mailto{hwitek@ffn.ub.es}
}

\begin{abstract}
A linear coupling between a scalar field and the Gauss--Bonnet invariant is the only known interaction term between a scalar and the metric that: respects shift symmetry; does not lead to higher order equations;  inevitably introduces black hole hair in asymptotically flat, 4-dimensional spacetimes. Here we focus on the simplest theory that includes such a term and we explore the dynamical formation  of scalar hair. In particular, we work in the decoupling limit that neglects the backreaction of the scalar onto the metric and evolve the scalar configuration numerically in the background of a Schwarzschild black hole and a collapsing dust star described by the Oppenheimer-Snyder solution. For all types of initial data that we consider, the scalar  relaxes at late times to the known, static, analytic configuration that is associated with a hairy, spherically symmetric black hole. This suggests that the corresponding black hole solutions are indeed  endpoints of collapse.
\end{abstract}

\maketitle

\section{Introduction}

	A century after black holes and gravitational waves were first predicted as solutions to Einstein's equations, the
	LIGO Scientific and VIRGO collaborations reported the first direct observations of gravitational waves
	originating from coalescing black-hole binaries~\cite{Abbott:2016blz,TheLIGOScientific:2016qqj,TheLIGOScientific:2016pea}.
	This remarkable discovery can also be considered as the first direct observation of black holes and has opened up an entirely new chapter
	in understanding and probing gravity in its strong-field regime \cite{TheLIGOScientific:2016src,Yunes:2016jcc}.
	It is possible that  electromagnetic and gravitational wave observations  of astrophysical black holes
	will reveal deviations from the (perturbed)  Kerr geometry and allow us to infer the existence of a new fundamental
	field~\cite{Barausse:2008xv, Berti:2015itd,Yagi:2016jml,Johannsen:2016uoh,Cardoso:2016ryw}.

	In general relativity, no-hair theorems have established that black holes are surprisingly simple objects, parametrized fully by only three (global) charges:
	their mass $M$, angular momentum $J$, and electromagnetic charge $Q$~\cite{Israel:1967wq,Israel:1967za,Carter:1971zc,Wald:1971iw,Bekenstein:1995un};
	see e.g. Refs.~\cite{Bekenstein:1996pn,Chrusciel:2012jk,Cardoso:2016ryw} for recent reviews on the topic.
	It is well-known that black holes can have hair
	in the presence of Yang-Mills fields~\cite{Volkov:1989fi,Bizon:1990sr,Greene:1992fw}.
	However,  our focus here will be extensions of general relativity that involve a scalar field. In such theories no-hair theorems still  exist. They are essentially a consequence of the fact that the equation
	\begin{equation}
	\label{eq:eq:scalareq}
		\Box \Phi
		=
		0\,,
	\end{equation}
	where $\Box$ is the curved spacetime d'Alembertian, admits only the trivial solution $\Phi = {\rm{constant}}$ in an asymptotically flat region of spacetime that has a Killing horizon as an inner boundary \cite{Hawking:1972qk}.
	This leads to the conclusion that stationary black hole solutions in scalar-tensor theories are the same as in general relativity.
	This result has been extended to scalars with nonlinear self-interactions in Ref.~\cite{Sotiriou:2011dz}. By means of field redefinitions and conformal transformations the applicability of the proof extends to the widest class of scalar-tensor theories that are quadratic in  derivatives.  Recent pedagogical reviews on no-hair theorems involving  scalar fields  can be found in Refs.~\cite{Sotiriou:2015pka,Herdeiro:2015waa}.

	As discussed in detail in Ref.~\cite{Sotiriou:2015pka}, no-hair theorems rely on a number of assumptions
	such as: asymptotic flatness and stationarity of the spacetime,
	absence of  matter,
	and the requirement that additional fields exhibit the same symmetries as the metric.
	The validity of these assumptions can be disputed. It is known that black holes can develop scalar hair if they have matter in their vicinity~\cite{Cardoso:2013opa,Cardoso:2013fwa},
	if the scalar is complex and has a time-dependent phase~\cite{Herdeiro:2014goa,
	Herdeiro:2016tmi,Herdeiro:2015gia}, or if the asymptotics are cosmological or anti-de Sitter \cite{Jacobson:1999vr,Horbatsch:2011ye,Torii:2001pg,Dias:2010ma,Dias:2011tj}.

	Still, the most obvious way to evade no-hair theorems is to consider a broader class of scalar-tensor theories in which the action contains terms with more than two derivatives.
	Horndeski~\cite{Horndeski:1974wa} has pinned-down
	the most general scalar-tensor theory that leads to second-order field equations. The action coincides with that of generalized galileons, which have recently received much attention in cosmology;
	see e.g. Ref.~\cite{Deffayet:2013lga} for a mathematical introduction and references therein for phenomenological applications.
	There is no no-hair theorem that applies to this general class of theories. Instead, a counter example has been known for quite some time. As shown in Ref.~\cite{Kanti:1995vq}, an exponential coupling between the scalar and the Gauss--Bonnet invariant $\G = R^{2} - 4~R_{a b}~R^{a b} + R_{a b c d}~R^{a b c d}$ leads to hairy black hole solutions. Such a coupling is known to be present
	 in the low energy effective action of heterotic string theory~\cite{Gross:1986mw,Metsaev:1987zx,Kanti:1995vq,Yunes:2013dva} and can also arise in a dimensional reduction of Lovelock gravity~\cite{Charmousis:2014mia,VanAcoleyen:2011mj}.
	Several studies of black holes in theories where a scalar is coupled to the Gauss--Bonnet invariant  have followed~\cite{Pani:2011gy,Yunes:2011we,Ayzenberg:2014aka,Maselli:2015tta,Kleihaus:2015aje}.

	An interesting subclass of Horndeski theories consists of the subset that satisfies shift symmetry, $\Phi \rightarrow \Phi + {\rm{constant}}$, as this symmetry prevents the scalar from acquiring a mass. This symmetry also excludes the exponential coupling $e^{\Phi} {\cal G}$ that led to the hairy solution of Ref.~\cite{Kanti:1995vq}.\footnote{More precisely, a coupling of the type $e^{\lambda \Phi} \G$ is invariant up to a redefinition of $\lambda$, but this {\em formal} invariance is not sufficient for the purposes of the proof presented in Ref.~\cite{Hui:2012qt}.}  Indeed,  it has been shown in Ref.~\cite{Hui:2012qt} that in shift-symmetric Horndeski theories  static, spherically symmetric, asymptotically flat black holes cannot have hair.
	However, as pointed out in Ref.~\cite{Sotiriou:2013qea}, there is a shift-symmetric coupling that manages to circumvent this no-hair theorem.  Since the Gauss--Bonnet invariant $\G$ is a total divergence, the linear coupling $\Phi \G$ is invariant under shifts up to a boundary term. Assuming it has a canonical kinetic term, the scalar satisfies the equation of motion
	\begin{equation}
	\label{eq:scalareq}
		\Box \Phi
		=
		- \lambda \G
		\,,
	\end{equation}
	where $\lambda$ is a coupling constant. Hence, the scalar is sourced by $\G$, which contains the Kretschmann scalar $R_{a b c d} R^{a b c d}$. Since the latter  in general does not vanish in a black hole spacetime, the scalar will be forced to have
	a nontrivial configuration. Note that another interesting way to circumvent the no-hair theorem of~Ref.~\cite{Hui:2012qt} is to allow $\Phi$ to have a linear dependence on Killing time \cite{Babichev:2013cya,Babichev:2016rlq,Sotiriou:2013qea}. However, we will not consider this option here.

	A solution to \eref{eq:scalareq} that describes the scalar profile of a hairy, static, spherically symmetric, black hole has been obtained in Ref.~\cite{Sotiriou:2013qea} working perturbatively in the coupling $\lambda$. This matches the solutions found earlier in Refs.~\cite{Campbell:1991kz,Yunes:2011we} using the same technique, but working with a more general coupling and applying a weak-field approximation for the scalar field as well. In Ref.~\cite{Sotiriou:2014pfa} instead, a nonperturbative, numerical solution has been presented and compared in detail with the perturbative one. This numerical solution resembles strongly the one found in Ref.~\cite{Kanti:1995vq} for the exponential coupling. All of these solutions are static and are expected to be  endpoints of gravitational collapse. Our main focus here is to present a first, preliminary exploration of whether this is indeed the case.

	Our motivation is threefold: (i) These solutions constitute a two-parameter family, parametrized by the mass and the scalar charge of the black hole. However, generically the scalar is singular on the event horizon, unless the scalar charge and the mass satisfy a bond. Imposing regularity selects a one-parameter family and it is an open question whether solutions within this family are dynamically selected during collapse. (ii) Stellar configurations in the theory in question have been shown to have vanishing scalar monopole, {\em i.e.}~the asymptotic fall-off for the scalar is necessarily faster than $r^{-1}$ \cite{Yagi:2015oca}. In contrast, in the known black hole solutions the scalar does exhibit an $r^{-1}$ fall-off. This implies that this monopolar component should develop during collapse. (iii) A scalar-tensor theory that evades no-hair theorems is expected to lead to detectable deviations from general relativity in the strong field regime. The first step towards confronting its prediction with observations is to understand black hole formation and evolution.

	Our exploration will be preliminary because we will resort to the decoupling limit, i.e.~we will neglect the scalar field's backreaction onto the geometry. This approximation reduces the problem to solving \eref{eq:scalareq} on a background spacetime that is a solution to Einstein's equations, potentially with matter. We will consider two different backgrounds: a Schwarzschild black hole, previously considered in Ref.~\cite{Benkel:2016kcq}, and an Oppenheimer-Snyder spacetime~\cite{Oppenheimer:1939ue},
	which
	is the simplest model
	of stellar collapse. The evolution of the scalar will correspond to the formation of scalar hair on these spacetimes.

	The rest of this paper is organized as follows. In Section \ref{sec:Setup} we define the theory with an action and derive equations of motion. We also discuss the decoupling limit and perturbative solutions. In Section \ref{sec:DynDecLimit} we formulate the problem in a way suitable for numerical methods and in Section \ref{sec:NumericalResults} we present our numerical results. Section \ref{sec:Conclusions} contains our conclusions.

\section{Setup}\label{sec:Setup}

	\subsection{Action and equations of motion}\label{ssec:ActEoM}

		The action that we will consider here reads
		\begin{equation}
		\label{eq:Action2Couplings}
			S
			=
			\int \dif^{4} x \sqrt{-g} \left[ \frac{R}{\kappa} + \mu \left(- \frac{1}{2} \nabla^{a} \Phi \nabla_{a}\Phi + \lambda  \Phi  \G \right) \right] + S_{\Psi}
			\,,
		\end{equation}
		where
		$\kappa = 16 \pi G$, $S_{\Psi}$ denotes the matter action, $\lambda$ and $\mu$ are coupling constants, and $\G$ is the Gauss--Bonnet invariant
		\begin{equation}
		\label{eq:DefGBInv}
			\G
			=
			R_{abcd}R^{abcd} - 4 R_{ab} R^{ab} + R^{2}
			\,.
		\end{equation}
		The coupling $\mu$ might appear redundant, as it could be absorbed in a redefinition of $\Phi$,
		but we choose to keep it for reasons that will become apparent in the next section. In the following we will employ geometric units $G=1$ and $c=1$.

		Varying the action with respect to the metric $g^{ab}$ and the scalar field $\Phi$ yields their field equations
		\begin{eqnarray}
			\label{eq:EoMsEdGBgeneralTen}
				G_{ab} + 16\pi\,\mu\,\lambda\,\G^{\rm{GB}}_{ab}
				=
				8\pi \left( T^{(\Psi)}_{ab} + \mu T^{(\Phi)}_{ab} \right)
				\,, \\
			\label{eq:EoMsEdGBgeneralSca}
				\Box \Phi
				=
				- \lambda \G
				\,.
		\end{eqnarray}
		If the matter stress-energy tensor
		\begin{equation}
			T^{(\Psi)}_{ab}
			\equiv
			\frac{2}{\sqrt{-g}}~\frac{\delta S_{\Psi}}{\delta g^{ab}}
			\,,
		\end{equation}
		is non-zero, as is the case for the collapsing, homogeneous dust star,
		we complement these by the conservation of the energy-momentum tensor and the continuity equation
		\begin{equation}
		\label{eq:EoMsMatter}
			\nabla^{b} T^{(\Psi)}_{ab}
			=
			0
			\,,\quad
			\nabla_{a}\left(E\, u^{a} \right)
			=
			0
			\,,
		\end{equation}
		where $E$ and $u^{a}$ are the rest-mass density and velocity of the dust.
		The canonical scalar field energy-momentum tensor is
		\begin{equation}
		\label{eq:TmnSF}
			T^{(\Phi)}_{ab}
			=
			\nabla_{a}\Phi \nabla_{b}\Phi - \frac{1}{2} g_{ab} \nabla^{c}\Phi \nabla_{c}\Phi
			\,,
		\end{equation}
		and the correction due to the Gauss--Bonnet term is
\begin{eqnarray}
\label{eq:TmnGB}
\G^{\rm{GB}}_{ab} & = &
	- 2 R \nabla^{}_{(a} \nabla^{}_{b)}\Phi
	- 4 R_{ab} \Box \Phi
	+ 4 R_{acbd} \nabla^{c} \nabla^{d}\Phi
\nonumber \\ & &
	+ 8 R_{c(a} \nabla^{c} \nabla^{}_{b)}\Phi
	+ 2 g_{ab} \left( R \Box\Phi - 2 R^{cd} \nabla_{c} \nabla_{d}\Phi \right)
\nonumber \\ & = &
	g^{}_{g(a} g^{}_{b)j} \epsilon^{ghcd} \epsilon^{ijef} R_{cdef} \nabla_{h} \nabla_{i}\Phi
\,.
\end{eqnarray}

		The action can be straightforwardly generalised by introducing a potential for $\Phi$, by generalising the coupling between $\Phi$ and $\G$, etc., but here we will focus on the simplest case that inevitably leads to hairy black holes. Any of these generalisations would break shift symmetry for $\Phi$.

	\subsection{The decoupling limit}

		We are interested in the dynamical development of scalar hair for black holes, so ideally we would like to study the evolution of the scalar field and its imprint on the spacetime geometry during gravitational collapse of a star and the formation of a black hole. However, for the sake of simplicity, we will consider a simpler problem, namely the evolution of the scalar field and hair formation in a given spacetime background.

This approximation can be formally derived from the original theory as a \textit{decoupling limit}. Consider  the  field equations \eref{eq:EoMsEdGBgeneralTen} and \eref{eq:EoMsEdGBgeneralSca}. By taking the limit $\mu \rightarrow 0$ one can turn off the backreaction of the scalar field on the metric and is left with
\begin{eqnarray}
\label{eq:EoMsGR}
	G_{ab} & = & 8\pi T^{(\Psi)}_{ab}
	\,, \\
\label{eq:EoMsEdGBdec}
	\Box \Phi & = & - \lambda \G
	\,,
\end{eqnarray}
together with \eref{eq:EoMsMatter}.
That is, the field equations for the metric reduce to  Einstein's equations in the presence of matter while the scalar's equation of motion remains unaffected.

\subsection{The nature and role of $\lambda$}

		It is important to stress that there are two distinct ways to view the theory~\eref{eq:Action2Couplings}  depending on the status of the coupling $\lambda$. If $\lambda$ is taken to be a usual coupling constant, the action can be taken as exact and studied as a classical
		theory of gravity. If $\lambda$ is instead considered to double as a book keeping parameter of an expansion,
		the action can be taken to describe some effective theory. In this case the theory is known to order $\lambda$ only, and hence one can only trust solutions to this order.

		To make this more concrete and rigorous let us define the dimensionless parameter $\varepsilon = \lambda / l^2$, where $l$ is a characteristic length scale,
		and consider the {\em small coupling limit} --- as opposed to decoupling --- where $\varepsilon \ll 1$. In the effective action scenario one has to work perturbatively in $\varepsilon$. Thus, the solutions will be of the form
\begin{eqnarray}
\label{eq:metric_perturbation}
g_{a b} & = & \overline{g}_{a b} + \varepsilon h_{a b} + \Ord{\varepsilon^2}{}
\,,\\
\label{eq:perturbative_expansion}
\Phi & = & \Phi_0 + \varepsilon \Phi_1 + \Ord{\varepsilon^2}{}
\,,
\end{eqnarray}
where the pair $(\overline{g}_{a b},\Phi_0)$ constitutes an exact, potentially dynamical solution of  the system \eref{eq:EoMsEdGBgeneralTen} and \eref{eq:EoMsEdGBgeneralSca} for $\varepsilon=0$ (or $\lambda=0$),
\begin{eqnarray}
\label{eq:EoMsEdGBgeneralTen0th}
G^{(0)}_{ab} = 8\pi \left( T^{(\Psi)}_{ab} + \mu T^{(0)}_{ab} \right)
\,, \\
\label{eq:EoMsEdGBOSgeneralSca0th}
\Box^{(0)} \Phi_0 = 0
\,.
\end{eqnarray}
		Here, $G^{(0)}_{ab}$, $T^{(0)}_{ab}$ and $\Box^{(0)}$ denote the Einstein tensor, the canonical scalar field
		energy-momentum tensor and the d'Alembertian constructed from the background fields $(\bar{g}_{ab},\Phi_{0})$.
		One can then use the expansion to generate a solution at  $\Ord{\varepsilon}{}$ by solving the equations
		\begin{eqnarray}
		\label{eq:EoMsEdGBOS1st}
			\label{eq:EoMsEdGBgeneralTen1st}
				G^{(1)}_{ab} + 16\pi\,\mu\, l^2 \G^{\rm{GB(0)}}_{ab}
				=
				8\pi  \mu T^{(1)}_{ab}
				\,,\\
			\label{eq:EoMsEdGBgeneralSca1st}
				\Box^{(0)} \Phi_1
				=
				- l^2 \G^{(0)}
				\,,
		\end{eqnarray}
		where quantities with superscript $^{(0)}$ are constructed from the background metric $\bar{g}_{ab}$,
		and $G^{(1)}_{ab}$ and $T^{(1)}_{ab}$ are the Einstein tensor and the scalar's stress tensor to first order. Higher order corrections should be discarded because the theory is only known to $\Ord{\varepsilon}{}$.

		Note that for this discussion $\mu$ has been taken to be ${\cal O}(1)$, as generically the decoupling limit has nothing to do with the small coupling limit we are discussing here. In fact, solutions with nontrivial $\Phi_0$ will have nonvanishing $T^{(0)}_{ab}$ and hence the scalar will have nonvanishing backreaction on the spacetime already at zeroth order in $\varepsilon$.
		Notably, stationary, asymptotically flat, black-hole spacetimes do have trivial $\Phi_0$.

		To see this one needs to first consider~\eref{eq:EoMsEdGBgeneralTen0th}~and~\eref{eq:EoMsEdGBOSgeneralSca0th}.
		These are effectively the equations of general relativity coupled to a scalar field. Hence, provided that the scalar shares the symmetries of the metric, no-hair theorems~\cite{Hawking:1972qk} apply and dictate that the only vacuum solution is $\Phi_0= {\rm{constant}}$ and the spacetime is described by the Kerr geometry. With $\Phi_{0}={\rm{constant}}$, \eref{eq:EoMsEdGBOS1st}~and~\eref{eq:EoMsEdGBgeneralSca1st} become exactly the same
		as~\eref{eq:EoMsGR}~and~\eref{eq:EoMsEdGBdec}.
		Hence, the {\em full} solution at decoupling will match
		the {\em leading order} solution at small coupling for stationary, asymptotically flat, black hole spacetimes.

		Another point we wish to clarify in this section is the role of $\lambda$ within the decoupling limit. Consider the transformation $\Phi \to \lambda \Phi$.
		At the level of the action~\eref{eq:Action2Couplings}, this transformation allows one to effectively set $\lambda$
		to $1$ by simply redefining $\mu$.
		This does not affect the process of taking the decoupling limit, and hence $\lambda$ becomes a redundant coupling at decoupling. The same can be seen at the level of the field equations. At decoupling $\mu\to 0$, $\lambda$ and $\Phi$ are entirely absent from \eref{eq:EoMsEdGBgeneralTen}. The transformation $\Phi \to \lambda \Phi$ makes $\lambda$ drop out from \eref{eq:EoMsEdGBgeneralSca} as well. Clearly, instead of generating solutions for different values of the coupling constant
		one can select a specific $\lambda$ and then obtain the remaining solutions simply by rescaling $\Phi$. Hence, from now on we will just set the dimensionless coupling
		$\lambda/M^{2}=1$.

	\subsection{The late-time behaviour of the scalar field}\label{ssec:AnalyticScalarSol}
	\label{latetime}

 The fact that the full solution at decoupling matches
		the  leading order solution at small coupling for any stationary, asymptotically flat, black hole spacetime is particularly relevant to our work. Static, spherically symmetric, asymptotically flat solutions to the theory in action \eref{eq:Action2Couplings} have been studied in Refs.~\cite{Sotiriou:2013qea,Sotiriou:2014pfa}. The small coupling solution is known analytically to quadratic order and it is unique. The leading order part of this solution, {\em i.e.}~the scalar configuration on a Schwarzschild background, will be the exact, static, asymptotically flat solution at decoupling.

		Below, we will use this scalar field profile to benchmark our numerical simulations in the decoupling
		limit at late times, when the field has settled down to a time-independent state. As explained in more detail in Section~\ref{ssec:BackgroundSpacetimes}, we numerically evolve the background spacetimes using puncture coordinates~\cite{Alcubierre:2002kk,Baker:2005vv,Campanelli:2005dd,vanMeter:2006vi} denoted as $(t,r,\theta,\phi)$.
At late times, these evolutions yield the well-known trumpet slices of the Schwarzschild
spacetime~\cite{Hannam:2006vv,Hannam:2008sg,Dennison:2014sma}.
However, because the metric functions in this slicing are not known in analytic form, here we instead employ isotropic
coordinates $(t_{\rm{S}},\rho,\theta,\phi)$.
The two coordinates systems agree within $\lesssim 0.1\%$ at late times and for radii $\rho \geq 10 M$ and $r \geq 10M$, as we have explicitly verified in~\ref{App:Coordinates}.
Hence, it is convenient to have the scalar profile of the analytically known static solution in isotropic coordinates. Instead of starting from the solution as given in Ref.~\cite{Sotiriou:2013qea} and perform a coordinate transformation, we prefer to rederive  the solution in the desired coordinate system.

		The Schwarzschild metric in isotropic coordinates is given by
		\begin{equation}
		\label{eq:SchwarzschildIsotropic}
			\dif s^{2}
			=
			- \alpha^{2}_{\rm{S}} \dif t^{2}_{\rm{S}} + \psi^{4} \left( \dif \rho^{2} + \rho^{2} \dif\Omega^{2} \right)
			\,,
		\end{equation}
		where the conformal factor and lapse function are
		\begin{equation}
		\label{eq:PsiAlpSchwarzschild}
			\psi
			=
			1 + \frac{M}{2\rho}
			\,,\quad
			\alpha^{2}_{\rm{S}}
			=
			\frac{ ( M-2\rho )^{2} }{ ( M+2\rho )^{2} }
			\,,
		\end{equation}
		and the horizon corresponds to $\rho_{\rm{H}} = M/2$.

		The scalar field equation~\eref{eq:EoMsEdGBdec}  then reads
		\begin{equation}
		\label{eq:ScaEoMDecSchwarzschild}
			\partial_{\rho\rho}\Phi(\rho)
			- \frac{8\rho}{M^{2}-4\rho^{2} } \partial_{\rho}\Phi(\rho)
			+ \lambda \frac{48 M^{2}}{\rho^{6} \psi^{8}(\rho)}
			=
			0
			\,.
		\end{equation}
		Direct integration leads to a solution with two integration constants. Fixing them by demanding
		(i) regularity at the horizon,
		and
		(ii) $\lim_{\rho\rightarrow\infty}\Phi = \Phi_{\infty}$
		yields
		\begin{equation}
		\label{eq:ScaSolSchwarzschild}
			\Phi(\rho)
			=
			\Phi_{\infty} + \frac{2\lambda}{3M \rho^{3} \psi^{6} } \left( 4 M^{2} + 3 \rho M \psi^{2} + 3 \rho^{2}\psi^{4} \right)
			\,.
		\end{equation}
		Due to shift symmetry we can always set $\Phi_{\infty}=0$ without loss of generality.
		This agrees with the solution given in Ref.~\cite{Sotiriou:2013qea}
		after applying the coordinate transformation $\bar{r}=\psi^2 \rho$, where $\bar{r}$ denotes the areal radius coordinate.

\section{Dynamics in the decoupling limit}
\label{sec:DynDecLimit}

	\subsection{Spacetime split revisited}
	\label{ssec:spacetimedecomposition}

		Since we plan to numerically evolve the system of field equations~\eref{eq:EoMsGR}~and~\eref{eq:EoMsEdGBdec}, we will perform the ADM-York decomposition
		common in numerical relativity~\cite{Arnowitt:1962hi,York1979,Alcubierre:2008,Cardoso:2014uka}.
		To this end we foliate the $4$-dimensional manifold $\left(\M,g_{ab}\right)$
		into a set of spatial hypersurfaces $\left(\Sigma_{t},\gamma_{ij}\right)$
		labelled by a time parameter $t$.
		We introduce the unit timelike vector $n^{a}$ orthogonal to the hypersurfaces
		with norm $n^{a}n_{a} = -1$.
		It can be expressed in terms of the lapse function $\alpha$ and shift vector $\beta^{i}$ as
		\begin{equation}
				\label{eq:NinAlpBet}
					n_{a}
					=
					-\alpha(1,0,0,0)
					\,,\quad
					n^{a}
					=
					\frac{1}{\alpha}\left(1,-\beta^{i}\right)
					\,.
		\end{equation}
		The $3$-metric $\gamma_{ab} = g_{ab} + n_{a} n_{b}$  acts as a projection
		operator
		\begin{equation}
		\label{eq:ProjOp}
			\gamma^{a}{}_{b}
			=
			\delta^{a}{}_{b} + n^{a} n_{b}
			\,,
		\end{equation}
		with $\gamma^{a}_{b} n^{b} = 0$ by construction.
		Then, the line element takes the form
		\begin{equation}
		\label{eq:LineElement3p1}
			\dif s^{2}
			=
			g_{ab} \dif x^{a} \dif x^{b}
			=
			- \left(\alpha^{2} - \beta^{k} \beta_{k} \right) \dif t^{2} + 2 \gamma_{ij} \beta^{i} \dif t \dif x^{j} + \gamma_{ij} \dif x^{i} \dif x^{j}
			\,.
		\end{equation}
		In the following we denote the covariant derivative and Riemann tensor with respect to the $3$-metric as $D_{i}$ and $R^{i}{}_{jkl}$,
		while the extrinsic curvature is
		\begin{equation}
		\label{eq:DefKij}
			K_{ij}
			=
			- \gamma^{c}{}_{i} \gamma^{d}{}_{j} \nabla_{c} n_{d}
			=
			- \frac{1}{2} \Lie_{n} \gamma_{ij}
			\,,
		\end{equation}
		where
		$\Lie_{n}$ is the Lie-derivative along $n^{a}$.

	\subsection{Background spacetime}\label{ssec:BackgroundSpacetimes}

		Dynamics  in the decoupling limit boil down to solving the scalar's equation  \eref{eq:EoMsEdGBdec} on a given background. In vacuum the Schwarzschild solution is the unique solution of \eref{eq:EoMsGR} under the assumption of spherical symmetry. Hence, without adding any  matter fields one can simply study the evolution of the scalar on the Schwarzschild geometry.
		Simple as this setup might be, it can still capture the most important aspects of the problem that we are trying to understand here, as one can still use it to model the formation of a nontrivial scalar configuration on a black hole spacetime. Moreover, one can check if the evolution indeed has the static solution discussed
in Sec.~\ref{ssec:AnalyticScalarSol}
as an endpoint. Hence, this will be the first case of background we will consider,
revisiting the results of Ref.~\cite{Benkel:2016kcq}.

		Once matter fields are included one needs to solve \eref{eq:EoMsGR} together with \eref{eq:EoMsMatter}.
		The solution of the system \eref{eq:EoMsMatter} and \eref{eq:EoMsGR} can be taken to represent a  star collapsing to form a black hole,
		while the solution of \eref{eq:EoMsEdGBdec} will represent the time evolution of $\Phi$ during and after the formation of the black hole.
		Here we will use the simplest spacetime that can be thought of as representing idealised stellar collapse: the Oppenheimer-Snyder solution~\cite{Oppenheimer:1939ue}.
		This is  an {\textit{analytic}} model of a homogeneous dust star collapsing into a black hole
		in spherical symmetry. Despite the fact that modelling matter as dust neglects important phenomena, such as the effect of radiation for instance, it seems to be an adequate approximation for our purposes. Since we are working in the decoupling limit, details regarding the structure of the matter configuration should not be particularly important for the behaviour of the scalar field, which is what is of interest here.

		In its exterior the Oppenheimer-Snyder collapse is described by the Schwarzschild solution
		\begin{equation}
		\label{eq:MetricOSExtSchwarzschild}
			\dif s^{2}
			=
			- f \dif t_{\rm{S}}^2 + \frac{1}{f} \dif \bar{r}^{2} + \bar{r}^{2} \dif\Omega^{2}
			\,,\,\,
			f(\bar{r})
			=
			1 - \frac{2M}{\bar{r}}
			\,,
		\end{equation}
in Schwarzschild coordinates $(t_{\rm{S}},\bar{r},\theta,\phi)$ with its surface located at $\bar{r}_{B}$.

The interior of the star is given by a closed Friedmann metric
\begin{equation}
\label{eq:MetricOSIntFriedmannTau}
\dif s^{2} = - \dif \tau^{2} + a^{2} \left( \dif\chi^{2} + \sin^{2}\chi \dif\Omega^{2} \right)
\,.
\end{equation}
Let us denote the surface of the star as $\chi_{B}$ in the coordinate system
$(\tau,\chi,\theta,\phi)$.
These can be related to the conformal time $\eta\in(-\pi,0)$ via
		\begin{equation}
		\label{eq:MetricOSIntFriedmannTauVsEta}
			\tau
			=
			a_{B} \left(\eta - \sin\eta \right)
			\,,\quad
			a
			=
			a_{B} \left(1 - \cos\eta \right)
			\,.
		\end{equation}
		Continuity of the metric is ensured by matching the circumference on the boundary between the
		star's interior and exterior regions, namely
		\begin{equation}
		\label{eq:OSRelRadii}
			a\,\sin\chi
			=
			\bar{r}
			=
			\psi^{2}\, \rho
			\,,
		\end{equation}
		where $\psi$ is the conformal factor and $\rho$ the isotropic radial coordinate.

		The initial scale factor $a_{B}$ and value of $\chi_{B}$
		can be expressed in terms of the areal radius $\bar{r}$
		\begin{equation}
			a^{2}_{B}
			=
			\frac{\bar{r}^{3}_{B}}{2M}
			\,,\quad
			\sin^{2}\chi_{B}
			=
			\frac{2 M}{\bar{r}_{B}}
			\,.
		\end{equation}

		Although both the Schwarzschild and the Oppenheimer-Snyder solution are known explicitly in the coordinate systems used above, these forms are not particularly suitable for our numerical simulations. In each case, one needs to introduce a foliation that penetrates the black hole horizon and at the same time allows us to continue the simulations after a black hole forms. To achieve this, instead of attempting to explicitly rewrite the solutions in a suitable foliation, we prefer to generate them numerically using $1+log$-slicing condition~\cite{Staley:2011ss}.  Details on the numerical evolution of the background spacetimes, including the prescription of initial data, can be found in \ref{App:BackgroundEvolution}.

	\subsection{Scalar field evolution}

		In order to evolve the scalar field equation \eref{eq:EoMsEdGBdec}
		in any of the two backgrounds we first  re-write it as a time-evolution problem.
		Therefore, we introduce the scalar's conjugate momentum
		\begin{equation}
		\label{eq:DefPi}
			\Pi
			=
			- \Lie_{n} \Phi.
		\end{equation}
		This definition immediately provides an evolution equation for the scalar field
		whereas the $3+1$-decomposition of \eref{eq:EoMsEdGBdec} yields
		the momentum's evolution. Hence, one has the set of equations
\begin{eqnarray}
\label{eq:EvolSFdecoupling}
(\partial_{t} - \Lie_{\beta}) \Phi & = & - \alpha \Pi
\,,\\
(\partial_{t} - \Lie_{\beta}) \Pi  & = &
        - \alpha \left(D^{i} D_{i} \Phi - K \Pi \right) - D^{i} \alpha D_{i} \Phi - \alpha\, \lambda \G
\,,\nonumber
\end{eqnarray}
		where
		$\Lie_{\beta}$ is the Lie-derivative along the shift vector,
		$D_{i}$ is the covariant derivative w.r.t. the  $3$-metric,
		and $K$ is the trace of the extrinsic curvature.
		Since we are working in the decoupling limit, the Gauss--Bonnet invariant $\G$ depends only on the background geometry.

		The system of evolution equations \eref{eq:EvolSFdecoupling} determines
		the scalar field dynamics in $3+1$ dimensions.
		They need to be supplemented with a set of initial conditions $(\Phi,\Pi)|_{t=0}$, and
		we will specify two different types.

		{\noindent{\bf{Initial Data 1:}}}
		The first set of data is for the trivial field configuration
		\begin{equation}
		\label{eq:SFID1}
			\Phi_{0}
			=
			0
			\,,\quad
			\Pi_{0}
			=
			0
			\,.
		\end{equation}
		This simple setup already leads to interesting results.
		In particular, it  demonstrates excellently that the scalar has to develop a nontrivial profile even if it is assumed to  be trivial initially,
		as it is sourced by the Gauss--Bonnet invariant.

		{\noindent{\bf{Initial Data 2:}}}
		The second type of initial data is a scalar field cloud anchored around the compact object
		given by
		\begin{equation}
		\label{eq:SFID2}
			\Phi_{0}
			=
			0
			\,,\quad
			\Pi_{0}
			=
			A_{0} \exp\left[\frac{(r-r_{0})^{2}}{\sigma^{2}} \right] \Sigma(\theta,\phi)
			\,,
		\end{equation}
		where $A_{0}$, $r_{0}$ and $\sigma$ are the amplitude, location and width of the Gaussian.
		$\Sigma(\theta,\phi)$ determines the angular distribution of $\Pi_0$ and is defined as a superposition of spherical harmonics.
		We focus on two specific choices, namely
		a spherically symmetric or ``monopole'' configuration with $\Sigma(\theta,\phi)=\Sigma_{00}\equiv Y_{00}$
		and a dipole configuration with $\Sigma(\theta,\phi)=\Sigma_{11}\equiv Y_{1-1} - Y_{11}$.
		Unless denoted otherwise we will always set the
		dimensionless amplitude $A_{0}/M=1$
		since it only leads to a re-scaling
		of the scalar in the decoupling limit.

\section{Numerical results}
\label{sec:NumericalResults}

	\subsection{Implementation}
	\label{ssec:CodeDescription}

We have implemented the field equations in the decoupling limit \eref{eq:EoMsGR}~and~\eref{eq:EoMsEdGBdec} as part of the
{\textsc{Lean}} code~\cite{Sperhake:2006cy}.
Originally based only on the {\textsc{Cactus}} Computational toolkit~\cite{Cactuscode:web,Goodale:2002a}
and the {\textsc{Carpet}} mesh refinement package~\cite{Schnetter:2003rb,CarpetCode:web},
{\textsc{Lean}} has now been adapted to the {\textsc{Einstein~Toolkit}}~\cite{EinsteinToolkit:web,Loffler:2011ay,Zilhao:2013hia}.
		We refer the interested reader to Ref.~\cite{ZilhaoWitek} for more details about the upgraded infrastructure.
		{\textsc{Lean}} has been extended to evolve additional bosonic fields coupled to gravity in
		Refs.~\cite{Witek:2012tr,Okawa:2014nda,Zilhao:2015tya}.

		To accomplish our present project we have not only
		incorporated new thorns into {\textsc{Lean}
		to evolve the  field equations in the decoupling limit but also new thorns capable of evolving the Oppenheimer-Snyder collapse.
		The details of this implementation are discussed in \ref{App:BackgroundEvolution}.
		We analytically prescribe
		initial data for the background spacetime and the scalar fields.
		We carry out simulations of the general relativity background using the $\chi$-version of the Baumgarte-Shapiro-Shibata-Nakamura (BSSN)
		formulation~\cite{Shibata:1995we,Baumgarte:1998te,Alcubierre:2008}
		together with puncture coordinates~\cite{Alcubierre:2002kk,Baker:2005vv,Campanelli:2005dd,vanMeter:2006vi}.
		We apply the method-of-lines to perform the evolutions,
		where spatial derivatives are typically approximated by fourth- or sixth-order finite difference stencils,
		and we use the fourth-order Runge-Kutta time integrator.
		In order to track the black-hole formation
		during the Oppenheimer-Snyder collapse and
		to obtain information about the black hole's
		properties we employ the apparent horizon finder
		{\textsc{AHFinderDirect}}~\cite{Thornburg:2003sf,Thornburg:1995cp}.
At the outer boundary we employ Sommerfeld, i.e. radiative, boundary conditions as implemented in the
{\textsc{Einstein Toolkit}}~\cite{EinsteinToolkit:web,Loffler:2011ay,Zilhao:2013hia}.

		Our numerical domain typically contains $7$ refinement levels, with the outer boundary located at $120M$ and
		resolution $h/M = 1.0$ on the outermost grid. This translates into the grid setup
		\begin{equation}
			\{(120, 24, 12, 6, 3, 1.5, 0.6),M/64\}
			\,,
		\end{equation}
		in the notation of Section IIE of Ref.~\cite{Sperhake:2006cy}, with resolution $h_{\rm{I}}/M=1/64$ on the innermost
		refinement level.
		While this setup is typically sufficent to obtain accurate numerical results, we found it
		necessary to push the outer boundary to $240M$ when we used Initial Data 2 and $\Pi_0$ was dipolar.

\begin{figure}
			\includegraphics[width=\textwidth]{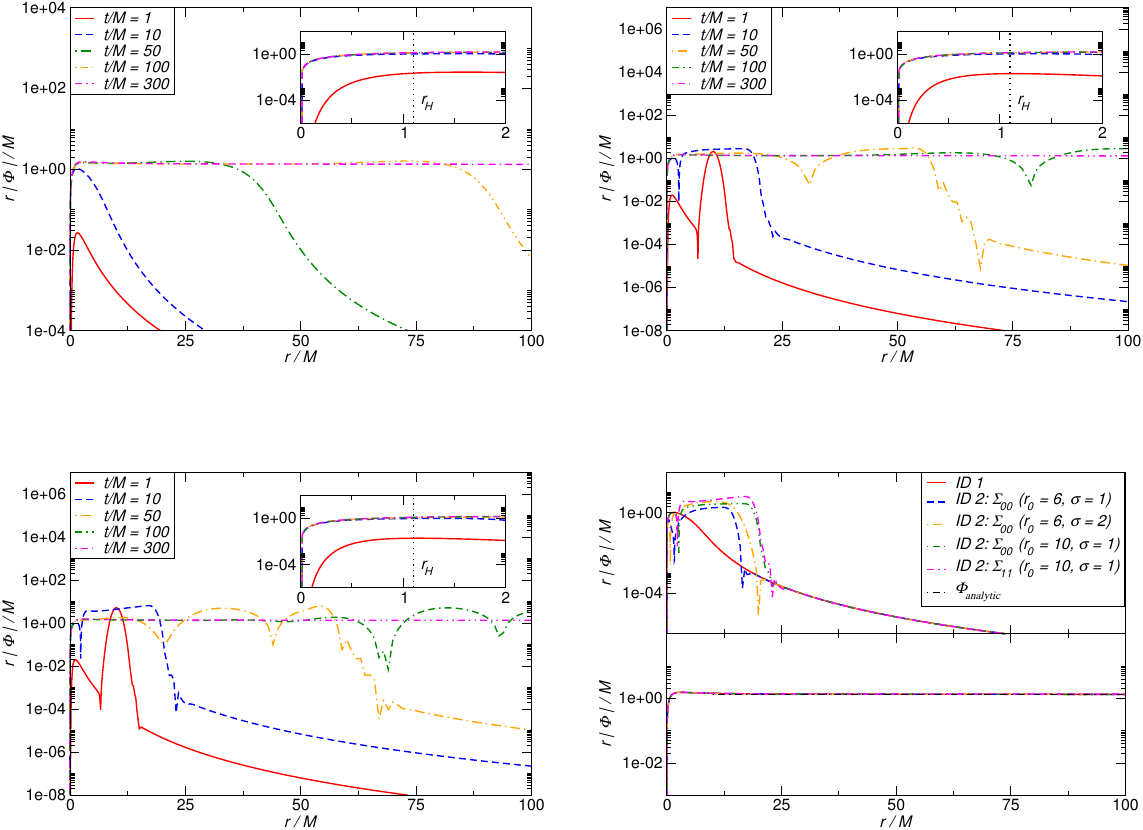}
			\caption{Radial scalar field profile, multiplied by the radius, at different instances of time in a Schwarzschild background for various initial data.
			{\em Top-left:} the field and its time derivative have been chosen to vanish initially  but the scalar still develops a nontrivial profile as it is sourced by the Kretschmann scalar.
			{\em Top-right:} the field vanishes initially and the derivative $\Pi_0$ is given as a spherically symmetric Gaussian shell with parameters
			$\Sigma(\theta,\phi)=\Sigma_{00}$, $r_{0}/M=10$ and $\sigma/M=1$ in \eref{eq:SFID2}.
			{\em Bottom-left:}  initially vanishing scalar with $\Pi_0$ given as a
			dipolar Gaussian shell with parameters
			$\Sigma(\theta,\phi)=\Sigma_{11}$, $r_{0}/M=10$ and $\sigma/M=1$ in \eref{eq:SFID2}.
			The  type of data breaks spherical symmetry. We present the profiles along the $\theta=0$ axis.
			During the evolution the scalar field sheds off its dipole moment through quasi-normal ringing
			as shown in the right panel of figure \ref{fig:Schwarzschild_Waveforms_Y11_ID2_Y11}
			and settles down to a spherical profile.
			{\em Bottom-right:}
			Comparison of early ($t/M=10$) and late time ($t/M=300$) profiles in Schwarzschild geometry for various initial configurations.
			In all cases, at late times the field converges to the known analytic
			solution \eref{eq:ScaSolSchwarzschild} with an asymptotic fall-off $r|\Phi|={\textrm{constant}}$, independently of the initial field content, as it is sourced by the Kretschmann scalar.
			}
		\label{fig:figuresSchw}
		\label{fig:Schwarzschild_ScalarRadialProfile_ID1}
		\label{fig:Schwarzschild_ScalarRadialProfile_ID2_Y00}
		\label{fig:Schwarzschild_ScalarRadialProfile_ID2_Y11}
		\label{fig:Schwarzschild_ScalarRadialProfile_DiffID}
		\end{figure}

		In order to estimate the numerical error we have performed benchmark tests against the analytic
		solution \eref{eq:ScaSolSchwarzschild} as well as convergence tests.
		The analysis, described in \ref{App:NumericalAccuracy} and illustrated in figures \ref{fig:NumErrorsSchwAndOS}
		and~\ref{fig:ConvergenceWaveforms},
		reveals a discretization error of $\Delta \Phi/ \Phi \lesssim 2\%$
		after an evolution time of about $t/M \sim 200$.
		At late times, the numerical solution agrees with the analytic one within less than
		$|\Phi/\Phi_{\rm{ana}} - 1|\lesssim 1\%$ for radii $r/M\geq10.0$.

		To analyze the formation of nontrivial scalar hair we consider both the field's radial profile
		as well as its multipolar components extracted on spheres of fixed radii $r_{\rm{ex}}$ as a function of time.
		In particular, we perform a multipole decomposition
		\begin{equation}
		\label{eq:MultipoleDecomposition}
			\Phi_{lm}(t,r_{\rm{ex}})
			=
			\int \dif\Omega\, \Phi(t,r_{\rm{ex}},\theta,\phi)\, Y^{\ast}_{lm}(\theta,\phi)
			\,,
		\end{equation}
		where $Y_{lm}(\theta,\phi)$ are the standard spherical harmonics.
		We also compute the  canonical scalar field energy density
		$E_{\rm{SF}} =  T^{(\Phi)}_{ab} n^{a} n^{b}$,
		where the energy-momentum tensor $T^{(\Phi)}_{ab}$ is given in \eref{eq:TmnSF}, and the quantity $E_{\rm{GB}} = \lambda \G^{\rm{GB}}_{ab} n^{a} n^{b}$, which can be interpreted as the contribution of the Gauss--Bonnet coupling to the scalar's total energy density. Though we have verified that both quantities remain finite and smooth in all cases, we do not present them or discuss them in  detail below. The scalar profile is always  smooth and these quantities do not provide any additional information in the decoupling limit.\footnote{Note that $E_{\rm{GB}}/E_{\rm{SF}} \sim\lambda$. We have argued that $\lambda$ is a redundant coupling within the decoupling limit approximation, but beyond decoupling the value of $\lambda$ will determine the relative importance of the two contributions in the backreaction the scalar will have on the metric and control potential deviations between decoupling and small coupling solutions.}

\begin{table}[b]
\centering
\caption{\label{tab:SetupSchwDec}
List of selected simulations performed in the background of a Schwarzschild black hole with mass parameter $M=1$
and  dimensionless coupling constant $\lambda/M^{2}=1$.
The scalar field is initially either trivial, i.e. Initial Data 1, or  $\Pi_0$ is given as a Gaussian shell
with angular distribution $\Sigma_{lm}$, located at $r_{0}/M$ and with width $\sigma/M$, i.e. Initial Data 2.
}
\vspace{0.5cm}
			\footnotesize
\begin{tabular}{l|lc}
\hline
Run                  & SF ID               & $(r_{0}/M,\sigma/M)$  \\
\hline
\verb|SBH_ID1|       & ID 1                & --              \\
\hline
\verb|SBH_Y00_r6w1|  & ID 2: $\Sigma_{00}$ & $(6.0,1.0)$     \\
\verb|SBH_Y00_r6w2|  & ID 2: $\Sigma_{00}$ & $(6.0,2.0)$     \\
\verb|SBH_Y00_r10w1| & ID 2: $\Sigma_{00}$ & $(10.0,1.0)$    \\
\verb|SBH_Y00_r10w2| & ID 2: $\Sigma_{00}$ & $(10.0,2.0)$    \\
\hline
\verb|SBH_Y11_r10w1| & ID 2: $\Sigma_{11}$ & $(10.0,1.0)$    \\
\hline
\end{tabular}
\end{table}

	\subsection{Scalar field dynamics around Schwarzschild black holes}
	\label{ssec:ResultsSchwarzschild}

		The specific choices of initial data for
		a scalar field in a Schwarzschild black hole background are summarized in table \ref{tab:SetupSchwDec} and include both Initial Data 1 and Initial Data 2 with a spherically symmetric or dipole scalar configuration.
		We illustrate the time evolution of the scalar field profile in detail for the various characteristic cases
		in figure \ref{fig:figuresSchw}.  All four panels actually present the radial profile rescaled by the radius, $r|\Phi|$,
		as this illustrates clearly the asymptotic behaviour. $r|\Phi|$ always remained smooth throughout the evolution.
		The presence of apparent kinks in the plots is due to the fact that we plot the absolute value of $\Phi$ and use a logarithmic scale.
		In all cases, the solutions approach $r |\Phi| = {\rm{constant}}$ for large radii at late times.
		This agrees well with the leading order behaviour
		$\Phi \sim \frac{2\lambda}{M r} + \mathcal{O}\left(\frac{1}{r^{2}}\right)$
		expected from the analytic solution \eref{eq:ScaSolSchwarzschild}.

		In fact, irrespective of the choice of initial data the solution always converges to the known static,
		analytic scalar profile of \eref{eq:ScaSolSchwarzschild} at late times.
		This can be seen in the the bottom-right panel of figure \ref{fig:Schwarzschild_ScalarRadialProfile_DiffID}
		where we show two time instances, one at early times and one at late times, for all the different initial data we considered. It can also be seen in more detail
		 in figure \ref{fig:NumErrorsSchwAndOS},
		where we show the fractional deviation between the static, analytic solutions and the late-time numerical profile for all the cases we studied.

		\begin{figure}[t]
			\centering
			\includegraphics[width=0.68\textwidth,clip]{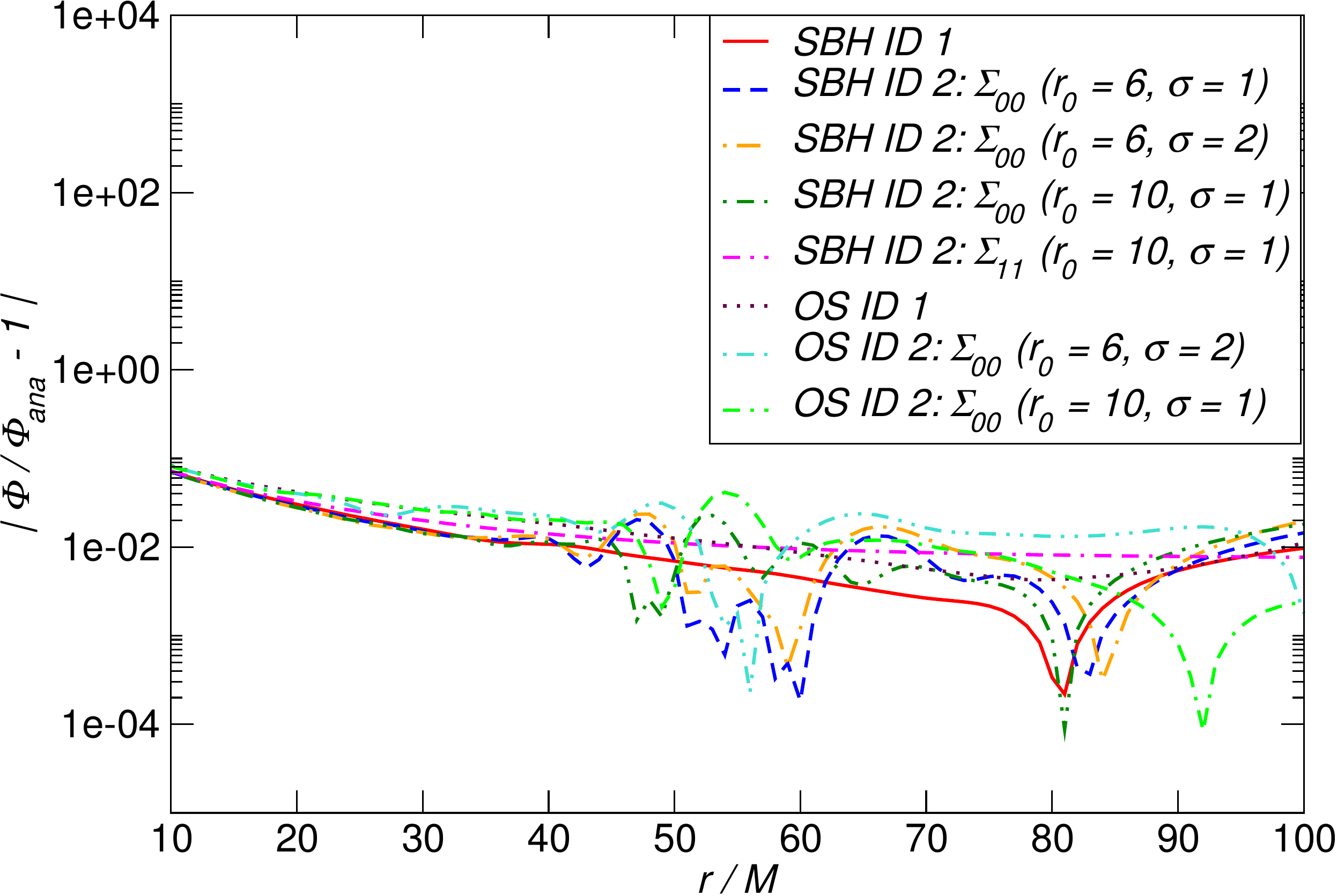}
			\caption{\label{fig:NumErrorsSchwAndOS}
			The fractional deviation between the late-time numerical profile  and the static, analytic solutions $|\Phi/\Phi_{\rm{ana}} - 1|$
			at late times $t/M=300$ for different types of initial data.
			We see that the deviation remains below $|\Phi/\Phi_{\rm{ana}} - 1| \lesssim 1.0\% $
			independent of the initial scalar field configuration.
``SBH'' refers to the Schwarzschild background whereas ``OS'' stands for the Oppenheimer-Snyder background.
			}
		\end{figure}

		\begin{figure}
			\includegraphics[width=\textwidth]{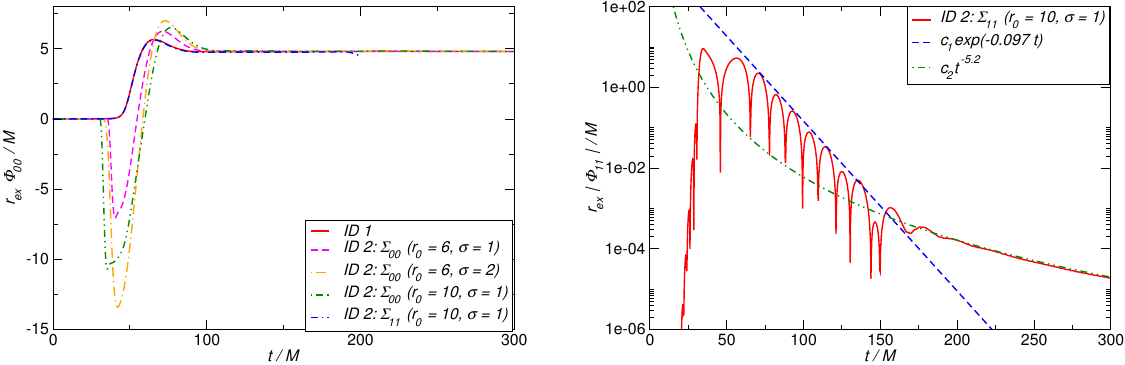}
			\caption{{\em Left:}
			$l=m=0$ multipole of the scalar field,
			re-scaled by the extraction radius $r_{\rm{ex}}/M=40$,
			evolved in the background of a Schwarzschild black hole.
			The different types of initial configurations,
			as indicated in the legend,
			determine the evolution at early times.
			Later on, after about $t/M\sim 100$ for this set of simulations, the scalar approaches the same solution independently
			of the initial data. {\em Right:} $l=m=1$ multipole of an initially dipole scalar configuration in a Schwarzschild geometry and
			re-scaled by the extraction radius $r_{\rm{ex}}/M=40$.
			The waveform clearly exhibits the quasi-normal ringdown with frequency $M\omega_{11}=0.292 - \imath0.097$,
			with the damping timescale indicated by the blue, dashed curve,
			as predicted by general relativity.
			This black-hole response is succeeded by a power-law fall-off. We have fitted the data to a function $\Phi\sim t^{-5.2}$
			(green dashed-dotted curve)
			which is in good agreement with predictions for the late-time tail in general relativity.}
		\label{fig:figuresWaveforms}
		\label{fig:Schwarzschild_Waveforms_Y00_DiffID}
		\label{fig:Schwarzschild_Waveforms_Y11_ID2_Y11}
		\end{figure}

		In the left panel of figure \ref{fig:Schwarzschild_Waveforms_Y00_DiffID} we illustrate the time evolution of the $l=m=0$ mode, constructed by projecting
		the scalar field onto the corresponding spherical harmonic as in \eref{eq:MultipoleDecomposition},
		and measured at a fixed coordinate radius $r_{\rm{ex}}/M=40$, for different initial configurations.
		The scalar field dynamics at early times are
		dominated by the specific initial setup,
		but  after $t/M\sim 100$ all types of data converge to the same solution.
		The case of an initially dipolar scalar field is of particular interest.
		The spherical component follows exactly that of an initially
		trivial field, while at the same time the scalar sheds off its dipolar component. This is depicted in the right panel of figure \ref{fig:Schwarzschild_Waveforms_Y11_ID2_Y11}
		where we present the $l=m=1$ multipole of the scalar field extracted at $r_{\rm{ex}}/M=40$.
		After the early time response, we clearly see the quasi-normal ringdown followed by the late-time tail.
		Moreover, we estimate the ringdown frequency of the numerical data to be $M\omega_{11} = 0.292 - \imath 0.097$.
		This is in excellent agreement, within $\lesssim 0.6\%$, with predictions in general relativity~\cite{Berti:2009kk}. The oscillatory ringdown phase is followed by a power-law decay $\Phi\sim t^{-5.2}$.
		This tail, computed from our time-domain data, agrees within $\lesssim 4\%$ with the theoretical prediction
		$\Phi\sim t^{-(2l + 3)} = t^{-5}$ for $l=1$ in general relativity~\cite{Price:1971fb,Leaver:1986gd,Ching:1995tj}.

	\subsection{Scalar field dynamics in Oppenheimer-Snyder background}\label{ssec:OSbackground}

		Next we evolve the scalar's field equation \eref{eq:EoMsEdGBdec} in the
		Oppenheimer-Snyder background.
		In practice, we evolve the
		\eref{eq:EoMsMatter} and
		\eref{eq:EoMsGR} in time as outlined in \ref{App:BackgroundEvolution}
		and with the initial data given in Section~\ref{ssec:BackgroundSpacetimes}.
		We set the intial size of the dust star to either $r_{\rm{B}}/M =5$ or $r_{\rm{B}}/M =10$, with the later resulting in a longer stellar phase.
		For practical purposes, we identify the time of collapse with the first appearance of an apparent horizon
		for which we have a time resolution of $\Delta t/M = 0.25$,
		and find $t_{\rm{AH}}/M = 21.25$ for $r_{\rm{B}}/M =5$
		and $t_{\rm{AH}}/M = 49$ for $r_{\rm{B}}/M =10$.
		We summarize our initial configurations in table \ref{tab:SetupOSDec}.

		\begin{figure}
			\includegraphics[width=\textwidth]{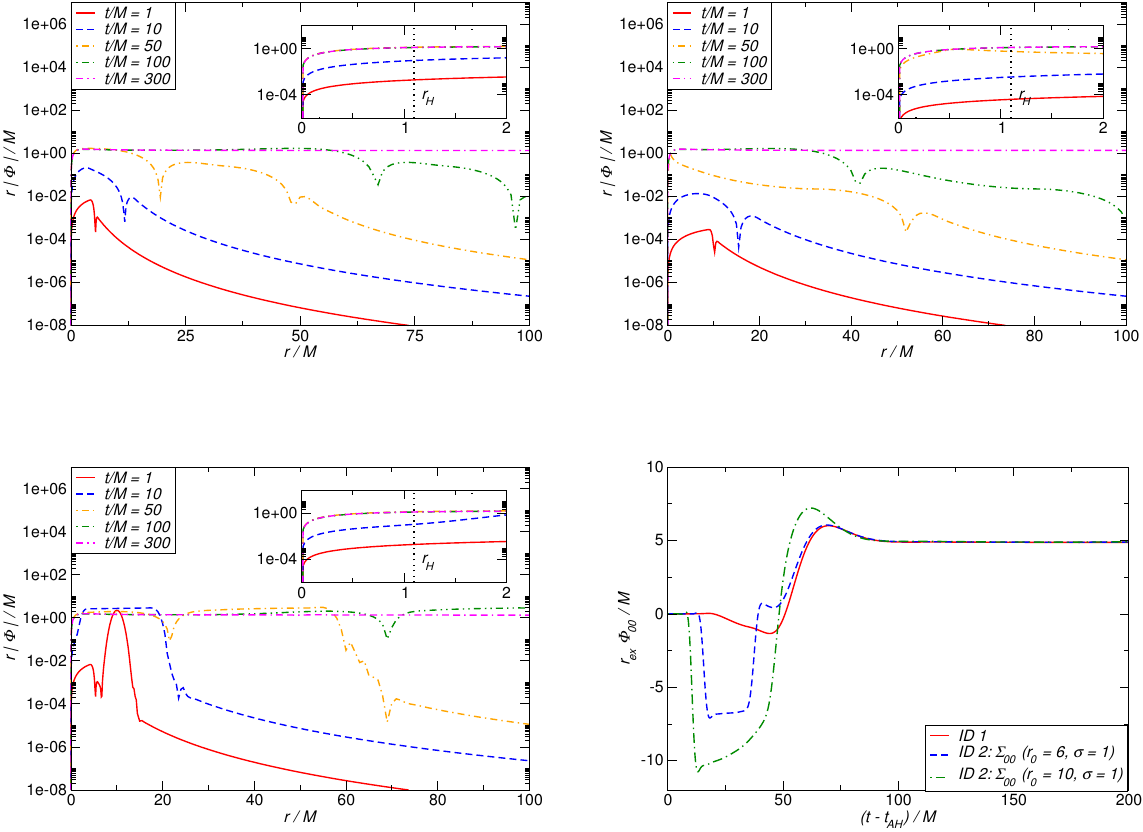}
			\caption{
			{\em Top-left:} Radial scalar field profile, multiplied by the radius, at different instances of time in an Oppenheimer-Snyder spacetime. The field and its time derivative have been chosen to vanish  initially and the  dust star has initial size $r_{\rm{B}}/M=5.0$.
			The apparent horizon forms after about $t_{\rm{AH}}/M\sim21.25$.
			{\em Top-right:} Same initial data
			but for a dust star of initial size $r_{\rm{B}}/M=10.0$. The apparent horizon forms after about  $t_{\rm{AH}}/M\sim49.0$.
			{\em Bottom-left:} Initial stellar configuration with $r_{\rm{B}}/M=5.0$ but
			different scalar field initial data, namely $\Pi_0$ given by a spherically symmetric Gaussian shell with parameters
			$\Sigma(\theta,\phi)=\Sigma_{00}$, $r_{0}/M=10$ and $\sigma/M=1$ in \eref{eq:SFID2}.
			{\em Bottom-right:} $l=m=0$ mode of the scalar field evolved in an Oppenheimer-Snyder geometry with $r_{\rm{B}}/M=5$
			for various initial configurations of the scalar.
			We have rescaled it by the extraction radius $r_{\rm{ex}}/M=40$ and shifted it in time by $t_{\rm{AH}}/M=21.25$
			signalling the black hole formation.
			The different types of initial configurations, as indicated in the legend,
			determine the evolution at early times.
			As can be seen in all figures, after the stellar collapse the scalar eventually approaches the
			known analytic solution in Schwarzschild spacetime \eref{eq:ScaSolSchwarzschild} and exhibits an  $r^{-1}$ asymptotic fall-off independent of the initial data.
			}
		\label{fig:figuresOS}
		\label{fig:OS_rB5_ScalarRadialProfile_ID1}
		\label{fig:OS_rB10_ScalarRadialProfile_ID1}
		\label{fig:OS_rB5_ScalarRadialProfile_ID2_Y00}
		\label{fig:OS_DiffrB_Waveforms_Y00_DiffID}
		\end{figure}

		\begin{table}[t]
		\centering
		\caption{\label{tab:SetupOSDec}
			List of selected simulations performed in the background of an Oppenheimer-Snyder dust collapse with
			initial surface radius $r_{B}/M$ and mass parameter $M=1$.
			The scalar field is initially either  trivial, i.e. Initial Data 1, or  $\Pi_0$ is given as a Gaussian shell
			with angular distribution $\Sigma_{lm}$, located at $r_{0}/M$ and with width $\sigma/M$, i.e. Initial Data 2.
			}
			\vspace{0.5cm}
			\footnotesize
			\begin{tabular}{l|clc}
				\hline
				Run                      & $r_{B}/M$ & SF ID               & $(r_{0}/M,\sigma/M)$ \\
				\hline
				\verb|OS_rB5_ID1|        & $5.0$     & ID 1                & --            \\
				\verb|OS_rB5_Y00_r6w1|   & $5.0$     & ID 2: $\Sigma_{00}$ & $(6.0,1.0)$   \\
				\verb|OS_rB5_Y00_r6w2|   & $5.0$     & ID 2: $\Sigma_{00}$ & $(6.0,2.0)$   \\
				\verb|OS_rB5_Y00_r10w1|  & $5.0$     & ID 2: $\Sigma_{00}$ & $(10.0,1.0)$  \\
				\hline
				\verb|OS_rB10_ID1|       & $10.0$    & ID 1                & --            \\
				\verb|OS_rB10_Y00_r6w1|  & $10.0$    & ID 2: $\Sigma_{00}$ & $(6.0,1.0)$   \\
				\hline
			\end{tabular}
		\end{table}

		In the top panels of figure \ref{fig:figuresOS} we present the radial profile multiplied by the radius, $r|\Phi|$, at different instances in time
		for a scalar that is initially entirely trivial in
		Oppenheimer-Snyder backgrounds with $r_{\rm{B}}/M=5$ and $r_{\rm{B}}/M=10$ respectively.
		For the bottom-left panel of figure \ref{fig:OS_rB5_ScalarRadialProfile_ID2_Y00} we have used Initial Data 2,
		with $\Pi_0$  being  a spherically symmetric Gaussian field in an Oppenheimer-Snyder background with $r_{\rm{B}}/M=5$.
		In all cases, the lines $t/M=1$ and $t/M=10$ correspond to the pre-collapse or stellar phase
		whereas the remaining curves signify the evolution in the resulting  black hole background.

		Already during the stellar phase the Gauss--Bonnet invariant forces the scalar to develop a nontrivial profile
		even if it is trivial initially.
		After a horizon forms and the exterior spacetime settles to a Schwarzschild black hole,
		we recover the behaviour already discussed in the previous section~\ref{ssec:ResultsSchwarzschild}:
		for all types of initial data the scalar approaches $r|\Phi| = {\rm{constant}}$ asymptotically.
		More specifically, as shown in figure \ref{fig:NumErrorsSchwAndOS},
		the entire configuration approaches the known, static, analytic solution in which the spacetime is described by
		a Schwarzschild black hole and the scalar profile is that of  \eref{eq:ScaSolSchwarzschild}.

		In the bottom-right panel of figure \ref{fig:OS_DiffrB_Waveforms_Y00_DiffID}  we present the $l=m=0$ multipole of the scalar
		for various field configurations evolved in the background of a collapsing dust star with
		initial radius $r_{\rm{B}}/M=5$.
		During the stellar phase, we clearly observe the excitation of a nontrivial scalar configuration,
		even in the case of trivial initial setup, induced by the Gauss--Bonnet invariant.
		After the collapse that occurs at around $t_{\rm{AH}}/M=21.25$ in this spacetime, however,
		the time evolution of the scalar becomes
		insensitive to the original geometry and exhibits the same behaviour as in Schwarzschild.
		In particular, the field again converges to the same hairy black hole solution
		regardless of the inital setup.

\section{Discussion}
\label{sec:Conclusions}

	We have investigated the dynamical formation of scalar hair in the simplest theory that fashions a linear coupling between a scalar field and the Gauss--Bonnet invariant.
	This coupling is known to yield black hole hair in stationary configurations.
	In order to  simplify our analysis we have worked in the decoupling limit, where the backreaction of the scalar onto the the spacetime geometry is neglected. This reduces the problem to solving
	the scalar's equations of motion in a background that is a solution to Einstein's equations.

We have considered two types of backgrounds, a Schwarzschild black hole
(see also Ref.~\cite{Benkel:2016kcq}) and the Oppenheimer-Snyder solution that describes the collapse of a dust star. We have explored several choices of initial data, including the  case of a trivial scalar with vanishing time derivatives, and nontrivial cases where the initial scalar configuration is
spherically symmetric or dipolar.
In all cases the scalar configuration eventually relaxes to the known, analytic, static configuration. Beyond decoupling this configuration corresponds to a hairy black hole. Although not a rigorous mathematical proof, this is a strong indication that this solution is indeed
the spherically symmetric endpoint of stellar collapse.

	It should be stressed that the known static configuration of \eref{eq:ScaSolSchwarzschild} is not the unique solution of \eref{eq:EoMsEdGBdec}. As discussed in Refs.~\cite{Sotiriou:2013qea,Sotiriou:2014pfa},  there exists a 2-parameter family of solutions  that generically diverge on the horizon. Imposing regularity on the horizon implies  a bond between the two parameters and
	selects a 1-parameter subclass. Although this appears to be a reasonable condition, one cannot know a priori if it constitutes tuning or if dynamical evolution naturally leads to this subclass. Our results imply the latter and, hence, clearly suggest that collapse will lead to the formation of hairy black holes in a theory where a scalar field couples linearly to the Gauss-Bonnet invariant.

	Among the cases of initial data we studied, the one where the time derivative of the scalar field is initially given by a dipolar Gaussian shell is of particular interest because it does not respect spherical symmetry. Since our backgrounds are spherically symmetric, within the decoupling approximation, the Gauss--Bonnet invariant fails to source
	non-spherical contributions. Hence, they decay with rates predicted by general relativity. For example, we have seen that a dipole field loses
	its dipole mode via quasi-normal ringing with frequencies matching those
	predicted in general relativity.
	At the same time, the field does develop a spherical profile that converges to the known, analytic, static solution.

	As is clear in our simulations, the dynamical behaviour of the scalar does not differ significantly when we switch from a Schwarzschild to an Oppenheimer-Snyder background. The early time behavior is affected, especially at small radii, and this can be attributed to the fact that the matter inside the star contributes to the curvature tensor, and effectively sources the scalar through the Gauss--Bonnet invariant. Our simulations clearly
	show that the scalar develops a nontrivial profile immediately during  the stellar phase and well before an apparent horizon forms. This is inevitable because it is sourced strongly by the Gauss--Bonnet invariant that does not vanish at any stage of the evolution.  It is worth pointing out that this is by no means in contradiction with the result of Ref.~\cite{Yagi:2015oca}, where it has been shown that stationary solutions of \eref{eq:EoMsEdGBdec} in an asymptotically flat spacetime without a horizon will have vanishing monopole.
	Firstly, our solutions are not stationary and when they approach stationarity at late times a horizon has already formed. Hence the result of Ref.~\cite{Yagi:2015oca} is not applicable here. Moreover, a vanishing monopole does not imply that the scalar configuration is trivial but only that its asymptotic fall-off is faster than $r^{-1}$ . Remarkably, this is indeed the case for our solutions before they reach stationarity.

	Our work has a number of exciting extensions. Within the decoupling limit, there are two natural next steps:  to use a more realistic stellar collapse model than Oppenheimer-Snyder and to  relax the symmetry assumptions of the background so as to allow for rotating black holes.
	Work in both directions is underway. It is also important to go beyond the decoupling approximation, as this would allow one to calculate the effect that the scalar field configuration has on the spacetime. As has been discussed in Ref.~\cite{Sotiriou:2014pfa} for the static black hole case, the metric configuration changes significantly in the interior of the horizon once the scalar field's backreaction is taken into account. In that case the singularity has finite area and black holes have a mimimum mass \cite{Sotiriou:2014pfa}. Hence, it would be interesting to explore the dynamical formation of black holes beyond the decoupling approximation.

\ack

	We would like to thank Leor Barack and Leo Stein for helpful discussions.
	The research leading to these results has received funding from the European
	Research Council under the European Union's Seventh Framework Programme
	(FP7/2007-2013) / ERC grant agreement n.~306425 `Challenging General
	Relativity'. HW acknowledges financial support by the European Union's H2020 research and innovation program under the Marie Sklodowska-Curie grant agreement No. BHstabNL-655360.
	Computations were performed on the {\textsc{minerva}} HPC Facility at the University of Nottingham through grant HPCA-01926-EFR, on the {\textsc{cosmos}} HPC Facility at the University of Cambridge
	operated on behalf of the STFC DiRAC HPC Facility and
	funded by the STFC grants ST/H008586/1, ST/K00333X/1 and ST/J005673/1 and on the MareNostrum supercomputer operated by the Barcelona Supercomputing Center
	and funded under Grant No. FI-2016-3-0006 `New frontiers in numerical general relativity'.

\appendix

\section{Numerical accuracy}
\label{App:NumericalAccuracy}

	In order to verify our numerical implementation and to access its numerical accuracy we have performed
	(i) a comparison between the numerical and isotropic coordinates,
	(ii) a convergence analysis of the scalar field at late times,
	and (iii) a benchmark test against the analytic solution
	for representative evolutions in the background of a Schwarzschild black hole
	and an Oppenheimer-Snyder collapse with initial radius $r_{\rm{B}}/M=5$.

\subsection{Verifying the coordinates}\label{App:Coordinates}
	As stated above, both for the Schwarzschild and the Oppenheimer-Snyder solutions,
	we have generated the background spacetime numerically. In our simulations we used puncture coordinates $(t,r,\theta,\phi)$, which are expected to resemble isotropic coordinates $(t_{\rm{S}},\rho,\theta,\phi)$ with high accuracy at late times and sufficiently large radii. A way to verify this is to compare directly the lapse function $\alpha$, the shift vector $\beta^{i}$, and the $3$-metric $\gamma_{ij}$ as obtained by our simulations at late times with the same components as one can read them off the Schwarzschild metric in isotropic coordinates. In all cases we found agreement to within $\lesssim 0.1\%$ for $\rho/M\geq10$ and $r/M\geq10$. An  illustration  for the case of the lapse is given in figure \ref{fig:BackgroundCoordinates}. This justifies using isotropic coordinates to perform the comparison between our numerical solutions and the known, static, analytic solution for the scalar profile.

	\begin{figure}[htpb!]
		\centering
		\includegraphics[width=0.68\textwidth,clip]{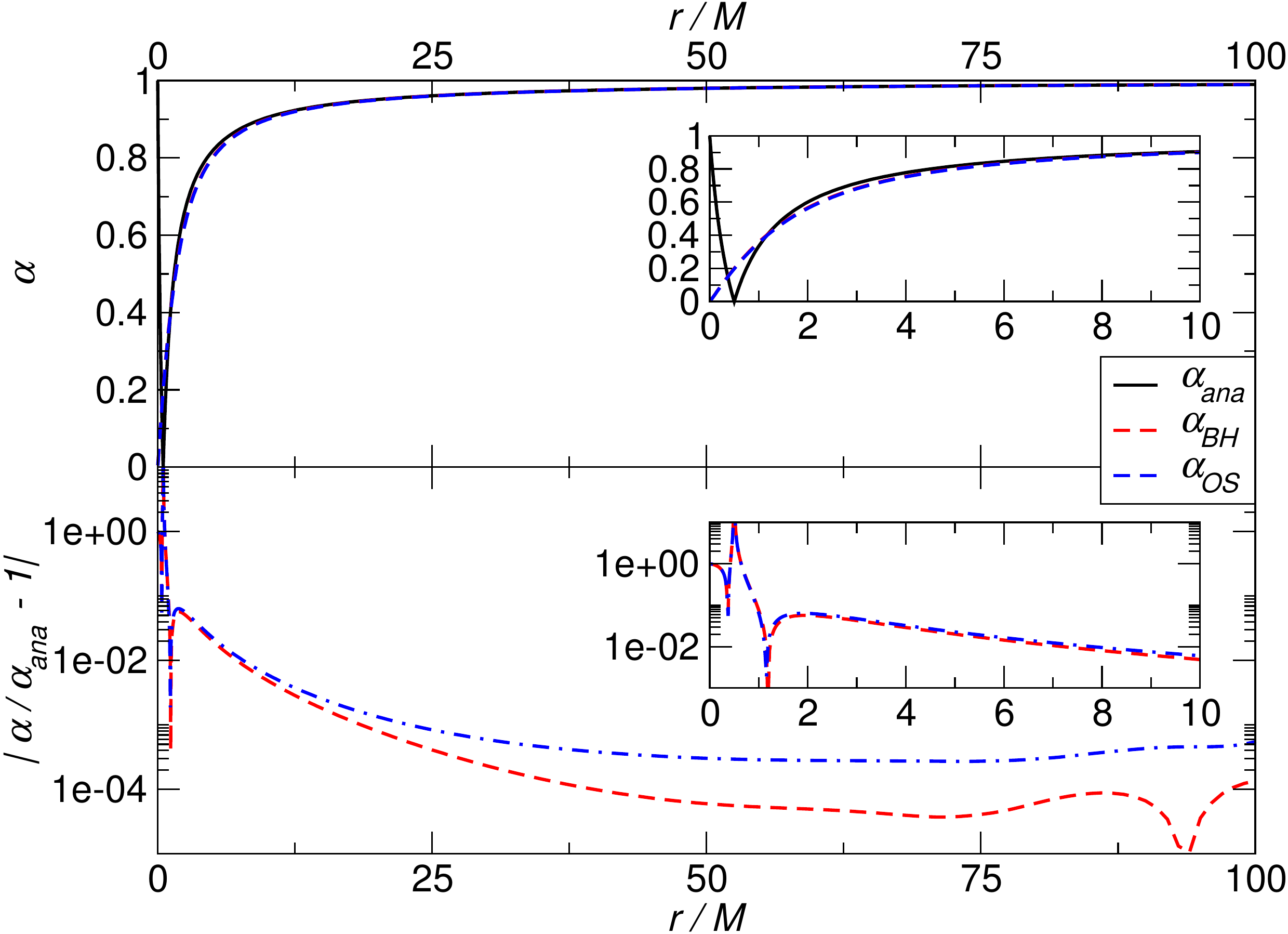}
		\caption{\label{fig:BackgroundCoordinates}
		{\em Top:}
		radial profile of the lapse function at $t/M=200$ for the numerically evolved
		Schwarzschild black hole (red dashed line) and Oppenheimer-Snyder collapse (blue dashed-dotted line)
		using the puncture gauge
		and its analytic value for a Schwarzschild black hole in isotropic coordinates (black solid line).
		{\em Bottom:}
		deviation between the numerically computed lapse function and its analytic value.
		As expected, the deviations are significant near the black hole and they drop below $0.1\%$ in the far region.
		Note that the axis label ``$r/M$'' really stands for both the (dimensionless) isotropic radial coordinate $\rho/M$
		and the (dimentsionless) puncture radial coordinate $r/M$.
		}
	\end{figure}

\subsection{Convergence analysis}
	We estimate the numerical error by performing a convergence analysis exemplarily for evolutions of Initial Data 1
	in \eref{eq:SFID1}
	in both types of background geometries.
	In particular, we have simulated this setup at three different resolutions
	$dx_{\rm{c}}/M=1.25$, $dx_{\rm{m}}/M=1.0$ and $dx_{\rm{f}}/M=0.75$
	of the outermost refinement level.

	We present the convergence plots for monopole mode $\Phi_{00}$ extracted at $r_{\rm{ex}}/M = 40$
	in figure \ref{fig:ConvergenceWaveforms}.
	Specifically, we show the difference between the coarse and medium, and medium and high resolution runs,
	where we have rescaled the latter by
	$Q_{4}=2.1$ in the Schwarzschild case and
	$Q_{2}=1.2$ for the Oppenheimer-Snyder evolution with $r_{\rm{B}}/M =5.0$
	indicating, respectively, $4^{\rm{th}}$ and $2^{\rm{nd}}$ order convergence.
	We estimate the numerical error in the scalar field
	to be about $\Delta \Phi_{00}/\Phi_{00} \lesssim 2\%$ after an evolution time of $t/M \sim 200$.

	\begin{figure}[htpb!]
		\centering
		\includegraphics[width=0.68\textwidth,clip]{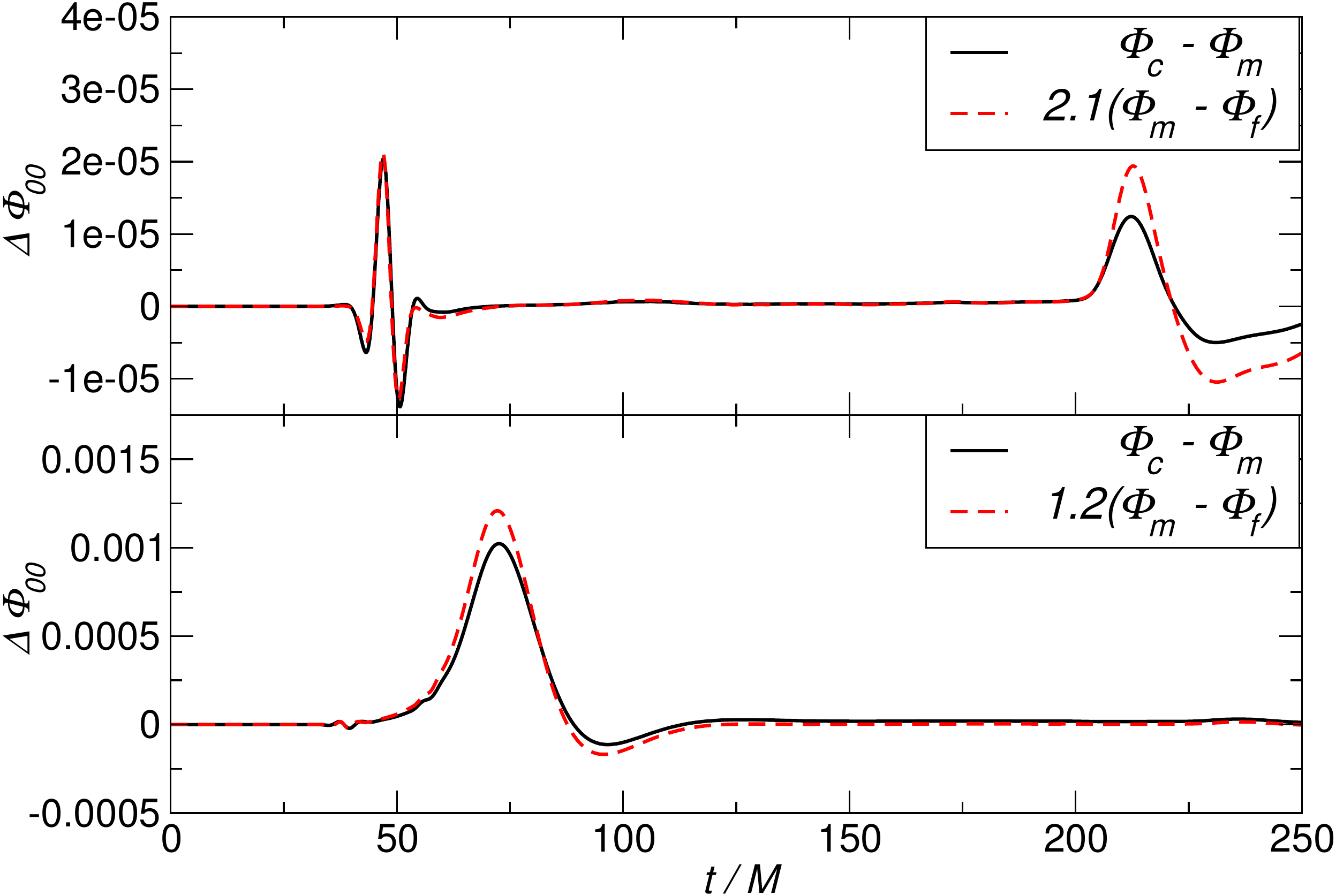}
		\caption{\label{fig:ConvergenceWaveforms}
		Convergence analysis of the scalar field initialized by Initial Data 1 and evolved in the background of a
		Schwarzschild black hole (top)
		and Oppenheimer-Snyder collapse (bottom), measured at $r_{\rm{ex}}/M = 40$.
		The rescaling of the difference between the medium and high resolution (red dashed lines) indicate, respectively,
		$4^{\rm{th}}$ and $2^{\rm{nd}}$ order convergence.
		}
	\end{figure}

	\subsection{Benchmark tests}
	As we have shown in the main body of the text, the scalar field numerically evolves towards the known, static, analytic solution given in \eref{eq:ScaSolSchwarzschild} for all cases we have studied.
	Considering this as the expected behaviour, we can  employ it to benchmark our numerical solution at late times.
	For this purpose, the fractional deviation between the numerical and analytic solutions  depicted in figure \ref{fig:NumErrorsSchwAndOS} can be reinterpreted as a relative error.
	For all cases, this relative error is
	$|\Phi/\Phi_{\rm{ana}} - 1|\lesssim 1.0\%$
	at late times $t/M=300$.

\section{Evolution of the background spacetime}
\label{App:BackgroundEvolution}

	In order to generate the background spacetimes numerically we employ a coordinate gauge that allows for a smooth evolution across the horizon, namely puncture coordinates~\cite{Alcubierre:2002kk,Baker:2005vv,Campanelli:2005dd,vanMeter:2006vi}.
	We use (standard) numerical relativity techniques that have been established over the last decade
	and details can be found, e.g., in Refs.~\cite{Shibata:1999hn,Alcubierre:2008,Staley:2011ss}.
	We will briefly summarize them here assuming the presence of dust, i.e., a homogeneous, pressure-less perfect fluid.
	We recover the black hole evolution for vanishing matter energy-momentum tensor and matter quantities.

	\subsection{Equations of motion}

		In the decoupling limit the equations of motion are given by \eref{eq:EoMsGR} together
		with \eref{eq:EoMsMatter} in the presence of matter, that is
		\begin{eqnarray}
		\label{eqApp:EoMGRConvervationContinuity}
			\label{eqApp:EoMGR}
				G_{ab}
				=
				8 \pi T^{(\Psi)}_{ab}
				\,, \\
			\label{eqApp:EoMConvervationContinuity}
				\nabla^{b} T^{(\Psi)}_{ab}
				=
				0
				\,, \quad
				\nabla_{a}( E\, u^{a})
				=
				0
				\,.
		\end{eqnarray}
		Here $G_{ab}=R_{ab}-1/2g_{ab}R$ is the Einstein tensor,
		the rest mass energy density $E$ vanishes in a black-hole background,
		and the energy-momentum tensor $T^{(\Psi)}_{ab}$ is
		\begin{equation}
		\label{eqApp:TabDust}
			T^{(\Psi)}_{ab}
			=
			\left\{\begin{array}{ll}
			        0 \\
			        E\, u_{a} u_{b}
			        \end{array} \right.
			        {\textrm{if}}\quad
			        \begin{array}{ll}
			        {\textrm{Schwarzschild}} \\
			        {\textrm{OS}}
			        \end{array}
			\,,
		\end{equation}
		where $E=E_{\rm{OS}}$ is the rest mass energy density of the collapsing dust star
		and $u^{a}$ its velocity field with normalization $u_{a} u^{a} = -1$.

	\subsection{Formulation as time-evolution problem}

		In order to numerically evolve the background spacetime it is convenient to perform a
		spacetime split as described in Sec.~\ref{ssec:spacetimedecomposition}.
		Recall, that
		we foliate the $4$-dimensional spacetime into a set of $3$-dimensional spatial hypersurfaces whose
		geometry is encoded in the $3$-metric $\gamma_{ij}$,
		its embedding is described by the extrinsic curvature $K_{ij}$ defined in \eref{eq:DefKij},
		and we introduce the unit normal vector $n^{a}$ orthogonal to the spatial slices.
		In the presence of dust we additionally decompose its velocity field according to
		\begin{equation}
		\label{eqApp:VelDust}
			u^{a}
			=
			w n^{a} + v^{a}
			\,,
		\end{equation}
		where $u^{a} n_{a} = -w$ and its spatial components are $v^{a}$ for which $v^{a} n_{a} = 0$ by construction.
		The normalization $u^{a}u_{a} = -1$ implies
		$w^{2} =  1 + v_{i} v^{i}$.
		It has proven convenient to redefine $E^{\ast}_{\rm{OS}} = w E_{\rm{OS}}$.

		Performing the ADM-York decomposition of the conservation and continuity
		equations \eref{eqApp:EoMConvervationContinuity}
		yields the evolution equations for the energy density and velocity field of the collapsing dust shell
\begin{eqnarray}
\label{eqApp:EvolDust}
\partial_{t} E^{\ast}_{\rm{OS}} & = &
        \Lie_{\beta}  E^{\ast}_{\rm{OS}}
        - \frac{E^{\ast}_{\rm{OS}}}{w} v^{i} D_{i} \alpha
        + \alpha K  E^{\ast}_{\rm{OS}}
\\ & &
        + \frac{\alpha}{w} \left[
        - D_{i}\left( E^{\ast}_{\rm{OS}} v^{i} \right)
        + E^{\ast}_{\rm{OS}} v^{i} D_{i} \left(\ln w \right)
        \right]
\,, \nonumber\\
\partial_{t} v_{i} & = &
        \Lie_{\beta} v_{i}
        - w D_{i} \alpha
        - \alpha \frac{1}{w} v^{j} D_{j} v_{i}
\,.
\end{eqnarray}
The $3+1$ split of Einstein's equations \eref{eqApp:EoMGR} yields the (gravity) evolution equations
\begin{eqnarray}
	\partial_{t} \gamma_{ij} & = & \Lie_{\beta} \gamma_{ij} - 2\alpha K_{ij}
	\,,\\
	\label{eqApp:EvolGRDust}
	\partial_{t} K_{ij} & = & \Lie_{\beta} K_{ij}
	        - D_{i} D_{j} \alpha
	        + \alpha \left( R_{ij} + K K_{ij} - 2 K_{ik} K^{k}{}_{j} \right)
	\nonumber \\ & &
	        - 4 \pi \alpha \frac{E^{\ast}_{\rm{OS}} }{w}
	          \left( 2 v_{i} v_{j} + \gamma_{ij} \right)
	\,,
\end{eqnarray}
and constraint equations
\begin{eqnarray}
\label{eqApp:ConstraintGRDustHamiltonian}
	\H & = & R - K^{ij} K_{ij} + K^{2} - 16\pi w E^{\ast}_{\rm{OS}}
	     = 0
	\,,\\
	\label{eqApp:ConstraintGRDustMomentum}
	\M_{i} & = & D^{j}K_{ij} - D_{i} K - 8 \pi E^{\ast}_{\rm{OS}} v_{i}
	         = 0
	\,.
\end{eqnarray}
		As indicated before, we recover those for a black hole by setting the energy density $E^{\ast}_{\rm{OS}}=0$.

	\subsection{Initial data}

		Let us first focus on the derivation of suitable initial data $(\gamma_{ij},K_{ij},\alpha,\beta^{i})|_{t=0}$
		that are complemented by the appropriate matter quantities in the case of the Oppenheimer-Snyder spacetime.
		To construct initial configurations of the background geometry, we need to solve the
		constraints~\eref{eqApp:ConstraintGRDustHamiltonian}~and~\eref{eqApp:ConstraintGRDustMomentum}.
		Therefore, we start by performing the York-Lichnerowicz conformal decomposition~\cite{Lichnerowicz1944,York:1971hw}
		of the metric
		\begin{equation}
		\label{eqApp:3metricAnsatz}
			\gamma_{ij}
			=
			\psi^{4} \hat{\gamma}_{ij}
			\,,\quad
			\hat{\gamma}_{ij}
			=
			{\rm{Diag}}\left[1,\rho^{2},\rho^{2}\sin^{2}\theta \right]
			\,,
		\end{equation}
		where $\psi$ and $\hat{\gamma}_{ij}$ are the conformal factor and metric.
		After applying this decomposition, the spatial line element becomes
		\begin{equation}
		\label{eqApp:MetricADMAnsatz}
			\dif l^{2}
			=
			\psi^{4} \left(\dif \rho^{2} + \rho^{2} \dif\Omega^{2} \right)
			\,.
		\end{equation}
		Comparing with \eref{eq:SchwarzschildIsotropic}
		we observe that this is nothing else but writing the initial spatial slices in isotropic coordinates
		$(\rho,\theta,\phi)$.

		Bearing in mind the definition of the extrinsic curvature \eref{eq:DefKij}
		we see immediately that $K_{ij}=0$ initially and, hence, the momentum constraint~\eref{eqApp:ConstraintGRDustMomentum}
		is satisfied trivially.
		Instead, the conformal factor will be specific to the particular spacetime under consideration
		and is constructed by solving the Hamiltonian constraint~\eref{eqApp:ConstraintGRDustHamiltonian}.
		Before we derive it for each of the cases below, let us provide the last piece of information to complete
		our (more generic) initial conditions,
		namely those for the gauge functions.

		Instead of taking the lapse function in isotropic coordinates,
		we initialize it either as $\alpha=1$ or as the pre-collapsed lapse,
		$\alpha = \psi^{-2}$,
		that has proven necessary for numerically stable simulations of black-hole spacetimes~\cite{Alcubierre:2008}.
		The shift vector is $\beta^{i}=0$.
		The gauge functions will adjust themselves to puncture coordinates by virtue of their evolution
		equations~\eref{eqApp:EvolGaugeLapse}~and~\eref{eqApp:EvolGaugeShift}.

		Let us now derive the conformal factor for each of the background spacetimes.

		\noindent{\em Schwarzschild solution:}
		The initial configuration is given by \eref{eqApp:3metricAnsatz} with the conformal factor
		\begin{equation}
		\label{eq:PsiSchwarzschild}
			\psi
			=
			1 + \frac{M}{2\rho}
			\,,
		\end{equation}
		and complemented with $K_{ij}=0$, $\alpha=\psi^{-2}$ and $\beta^{i}=0$.

		\noindent{\em Oppenheimer-Snyder solution:}
		To construct its initial configuration we write the spatial metric in the form \eref{eqApp:MetricADMAnsatz}.
		By using the matching conditions \eref{eq:OSRelRadii}
		we find the coordinate transformations
		\begin{equation}
			\bar{r}
			=
			\rho \left( 1+\frac{M}{2\rho} \right)^{2}
			\,,\quad
			\sin\chi
			=
			\frac{ 2\rho \sqrt{2M\rho^{3}_{\rm{B}}} }{ 2\rho^{3}_{\rm{B}} + M \rho^{2} }
			\,,
		\end{equation}
		where $\rho_{\rm{B}}$ denotes the surface radius of the dust star in isotropic coordinates.
		Then the initial state of the Oppenheimer-Snyder collapse is prescribed by \eref{eqApp:3metricAnsatz}
		with
		\begin{equation}
		\label{eq:OSInitDataPsi}
			\psi
			=
			\left\{\begin{array}{ll}
			              1+\frac{M}{2\rho} \\
			              \left[ \frac{( M+2\rho_{\rm{B}} )^{3}}{4 ( 2\rho^{3}_{B} + M \rho^{2} )} \right]^{1/2}
			              \end{array} \right.
			        {\textrm{if}}\quad
			        \begin{array}{ll}
			        \rho > \rho_{\rm{B}} \\
			        \rho \leq \rho_{\rm{B}}
			        \end{array}
			\,,
		\end{equation}
		and complemented by
		$K_{ij}=0$, $\alpha=1$, $\beta^{i}=0$ and
		\begin{equation}
		\label{eq:OSInitDataRho}
			E_{\rm{OS}}
			=
			\left\{\begin{array}{ll}
			        0  \\
			        \frac{48}{\pi} \frac{M \rho^{3}_{B}}{ ( M + 2 \rho_{\rm{B}} )^{6} }
			        \end{array}\right.
			        {\textrm{if}}\quad
			        \begin{array}{ll}
			        \rho > \rho_{\rm{B}} \\
			        \rho \leq \rho_{\rm{B}}
			        \end{array}
			\,.
		\end{equation}

	\subsection{Evolution equations}

	\begin{figure}[t!]
		\centering
		\includegraphics[width=0.68\textwidth,clip]{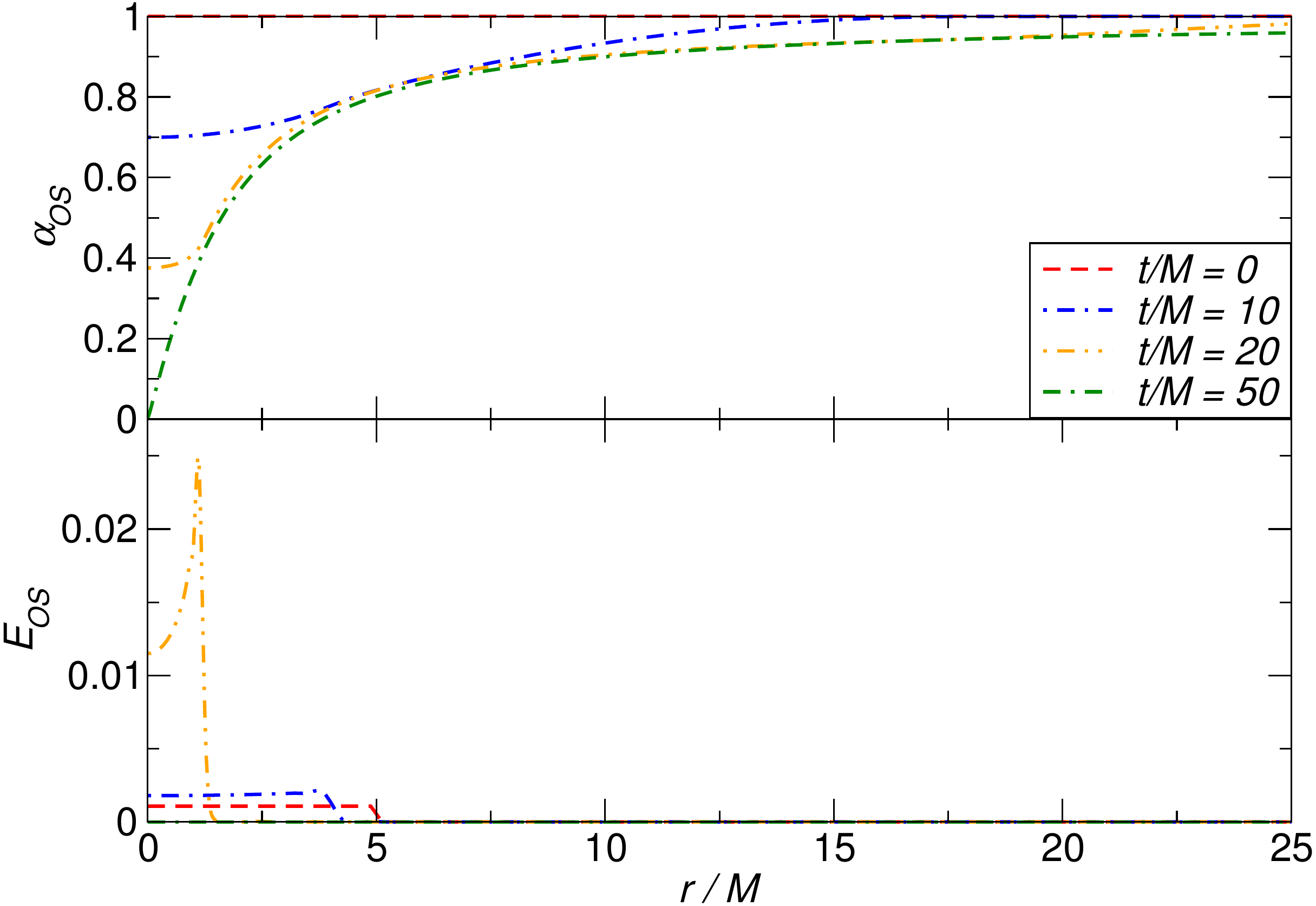}
		\caption{\label{fig:OScollapse}
		Evolution of the Oppenheimer-Snyder background with initial radius $r_{\rm{B}}/M=5$,
		corresponding to an areal radius of $\bar{r}_{\rm{B}}/M = 6.05$.
		We present the radial profiles of the lapse function (top) and energy density (bottom)
		at different instances in time.
		The orange line correspond approximately to the time of collapse; the apparent horizon first formed at
		$t_{\rm{AH}}/M\sim21.25$.
		}
		\end{figure}

		To follow the time development of the background numerically we adopt a free-evolution scheme, i.e.,
		we solve the constraints~\eref{eqApp:ConstraintGRDustHamiltonian}~and~\eref{eqApp:ConstraintGRDustMomentum} only for the initial data,
		which is then evolved.
		Throughout the evolution we monitor the constraints and verify that they remain satisfied within the numerical accuracy.
		In practice, we evolve~\eref{eqApp:EvolDust}~--~\eref{eqApp:EvolGRDust} using the BSSN formulation of Einstein's equations~\cite{Shibata:1995we,Baumgarte:1998te} which is known to yield numerically stable evolutions.
		In this approach, the dynamical variables are given by
		\begin{eqnarray}
		\label{eqApp:BSSNvars}
			\chi = \gamma^{-1/3}
			\,,\quad
			\tilde{\gamma}_{ij} = \chi \gamma_{ij}
			\,,\quad
			\tilde{\Gamma}^{i} = \tilde{\gamma}^{jk} \tilde{\Gamma}^{i}{}_{jk}
			\,,\nonumber \\
			K = \gamma^{ij} K_{ij}
			\,,\quad
			\tilde{A}_{ij} = \chi \left( K_{ij} - \frac{1}{3}\gamma_{ij} K \right)
			\,,\nonumber \\
			E^{\ast}_{\rm{OS}}  = E^{\ast}_{\rm{OS}}
			\,,\quad
			\tilde{v}_{i} = v_{i}
			\,,\quad
			\tilde{v}^{i} = \frac{1}{\chi} v^{i}
			\,. \nonumber
		\end{eqnarray}
		The evolution equations are further modified by appropriate constraint addition,
		and their explicit form can be found, e.g., in Ref.~\cite{Alcubierre:2008}.

		The system of evolution PDEs~\eref{eqApp:EvolDust}~--~\eref{eqApp:EvolGRDust}
		is closed by a suitable choice of coordinate conditions.
		In particular, we employ puncture coordinates~\cite{Alcubierre:2002kk,Baker:2005vv,Campanelli:2005dd,vanMeter:2006vi}
		\begin{eqnarray}
		\label{eqApp:EvolGaugeLapse}
		\partial_{t} \alpha & = & \beta^{k} \partial_{k} \alpha - 2 \alpha K
		\,,\\
		\label{eqApp:EvolGaugeShift}
		\partial_{t} \beta^{i} & = & \beta^{k} \partial_{k} \beta^{i} + \zeta_{\Gamma} \tilde{\Gamma}^{i} - \eta_{\beta} \beta^{i}
		\,,
		\end{eqnarray}
		where we set the parameters to $\eta_{\beta} = 1/M$ and $\zeta_{\Gamma} = 3/4$.

		We illustrate the evolution of the Oppenheimer-Snyder collapse in figure \ref{fig:OScollapse}
		where we depict the lapse function and energy density at different instances in time.
		The results are in good agreement with those presented in Ref.~\cite{Staley:2011ss}.

\bibliographystyle{iopart-num}
\bibliography{hairformationGB_v3.bib}

\end{document}